\pdfoutput=1

\documentclass[11pt,twoside,a4paper,cmspaper,final,collab]{cms-tdr}
\def\svnVersion{477336P}\def\cmsCernNoTag{CERN-EP-2018-094}\def\cmsCernDate{\today}\def\cmsMessage{Published in Physical Review Letters as \href{http://dx.doi.org/10.1103/PhysRevLett.121.141802}{\doi{10.1103/PhysRevLett.121.141802.}}}

\begin{document}\cmsNoteHeader{EXO-17-022}

\hyphenation{had-ron-i-za-tion}
\hyphenation{cal-or-i-me-ter}
\hyphenation{de-vices}
\RCS$HeadURL: svn+ssh://svn.cern.ch/reps/tdr2/papers/EXO-17-022/trunk/EXO-17-022.tex $
\RCS$Id: EXO-17-022.tex 472127 2018-08-17 14:29:01Z johnpaul $

\newlength\cmsTabSkip\setlength{\cmsTabSkip}{1ex}
\newlength\cmsFigWidth
\ifthenelse{\boolean{cms@external}}{\setlength\cmsFigWidth{0.48\columnwidth}}{\setlength\cmsFigWidth{0.40\textwidth}}
\ifthenelse{\boolean{cms@external}}{\providecommand{\cmsLeft}{top\xspace}}{\providecommand{\cmsLeft}{left\xspace}}
\ifthenelse{\boolean{cms@external}}{\providecommand{\cmsRight}{bottom\xspace}}{\providecommand{\cmsRight}{right\xspace}}
\ifthenelse{\boolean{cms@external}}{\providecommand{\cmsLeftCap}{Top\xspace}}{\providecommand{\cmsLeftCap}{Left\xspace}}
\ifthenelse{\boolean{cms@external}}{\providecommand{\cmsRightCap}{Bottom\xspace}}{\providecommand{\cmsRightCap}{Right\xspace}}
\ifthenelse{\boolean{cms@external}}{\providecommand{\NA}{\ensuremath{\cdots}\xspace}}{\providecommand{\NA}{\ensuremath{\text{---}}\xspace}}
\ifthenelse{\boolean{cms@external}}{\providecommand{\CL}{C.L.\xspace}}{\providecommand{\CL}{CL\xspace}}
\ifthenelse{\boolean{cms@external}}{\providecommand{\CLndot}{C.L}}{\providecommand{\CLndot}{CL}}
\ifthenelse{\boolean{cms@external}}{\providecommand{\suppMaterial}{the supplemental material~\cite{bib:suppMaterial}\xspace}} {\providecommand{\suppMaterial}{Appendix~\ref{app:suppMat}\xspace}}
\ifthenelse{\boolean{cms@external}}{\providecommand{\suppMaterialRef}[2]{#1}}{\providecommand{\suppMaterialRef}[2]{#2}}

\newcommand{\mavg}{\ensuremath{\overline{m}}\xspace}
\newcommand{\masy}{\ensuremath{A_\mathrm{m}}\xspace}
\newcommand{\deta}{\ensuremath{\Delta\eta}\xspace}
\newcommand{\tauN}{\ensuremath{\tau_N}\xspace}
\newcommand{\tautwoone}{\ensuremath{\tau_{21}}\xspace}
\newcommand{\taufourtwo}{\ensuremath{\tau_{42}}\xspace}
\newcommand{\taufourthree}{\ensuremath{\tau_{43}}\xspace}
\newcommand{\taukl}{\ensuremath{\tau_{kl}}\xspace}
\newcommand{\pdfm}{\ensuremath{P(m)}\xspace}
\newcommand{\pdfmavg}{\ensuremath{P_{\text{avg}}(\overline{m})}\xspace}
\newcommand{\squarkoff}{\ensuremath{\widetilde{\cmsSymbolFace{q}}^*}\xspace}
\newcommand{\higgsino}{\ensuremath{\widetilde{\cmsSymbolFace{H}}}\xspace}

\cmsNoteHeader{EXO-17-022}
\title{\texorpdfstring{
	Search for pair-produced resonances each decaying into at least four quarks in proton-proton collisions at $\sqrt{s}=13\TeV$
	}{
	Search for pair-produced resonances each decaying into at least four quarks in proton-proton collisions at sqrt[s] = 13 TeV
}}

\date{\today}

\abstract{
This letter presents the results of a search for pair-produced particles of masses above 100\GeV that each decay into at least four quarks.
Using data collected by the CMS experiment at the LHC in 2015-2016, corresponding to an integrated luminosity of 38.2\fbinv, reconstructed particles are clustered into two large jets of similar mass, each consistent with four-parton substructure. No statistically significant excess of data over the background prediction is observed in the distribution of average jet mass. Pair-produced squarks with dominant hadronic $R$-parity-violating decays into four quarks and with masses between 0.10 and 0.72\TeV are excluded at 95\% confidence level. Similarly, pair-produced gluinos that decay into five quarks are also excluded with masses between 0.10 and 1.41\TeV at 95\% confidence level. These are the first constraints that have been placed on pair-produced particles with masses below 400\GeV that decay into four or five quarks, bridging a significant gap in the coverage of $R$-parity-violating supersymmetry parameter space.
}

\hypersetup{%
pdfauthor={CMS Collaboration},%
pdftitle={Search for pair-produced resonances each decaying into at least four quarks in proton-proton collisions at sqrt(s) = 13 TeV},%
pdfsubject={CMS},%
pdfkeywords={CMS, physics, resonances}}

\maketitle

Despite the many successes of the CERN LHC program, a solution to the hierarchy problem~\cite{witten,dimopoulos}, which arises from the large difference between the electroweak and Planck scales, remains elusive. One theoretically attractive resolution is supersymmetry (SUSY)~\cite{golfand-1971,wess-1974,nilles-1984,barbieri-1982,Dimopoulos:1981yj,Ibanez:1981yh,Marciano:1981un}.
Arguments based on naturalness require that SUSY particles---specifically gluinos (\PSg), third-generation squarks (\PSQ), and higgsinos (\higgsino)---be accessible at the current LHC energies~\cite{BARBIERI198863,Kitano:2006gv,Brust:2011tb,Papucci2012}.  While $R$-parity-conserving models of SUSY~\cite{rpc} have been emphasized because they can parsimoniously accommodate a dark matter candidate, $R$-parity-violating (RPV) models still lack constraints in a number of important scenarios~\cite{Evans2013,RPVModel}.

\begin{figure}[htbp]
	\centering
	\includegraphics[width=0.95\cmsFigWidth]{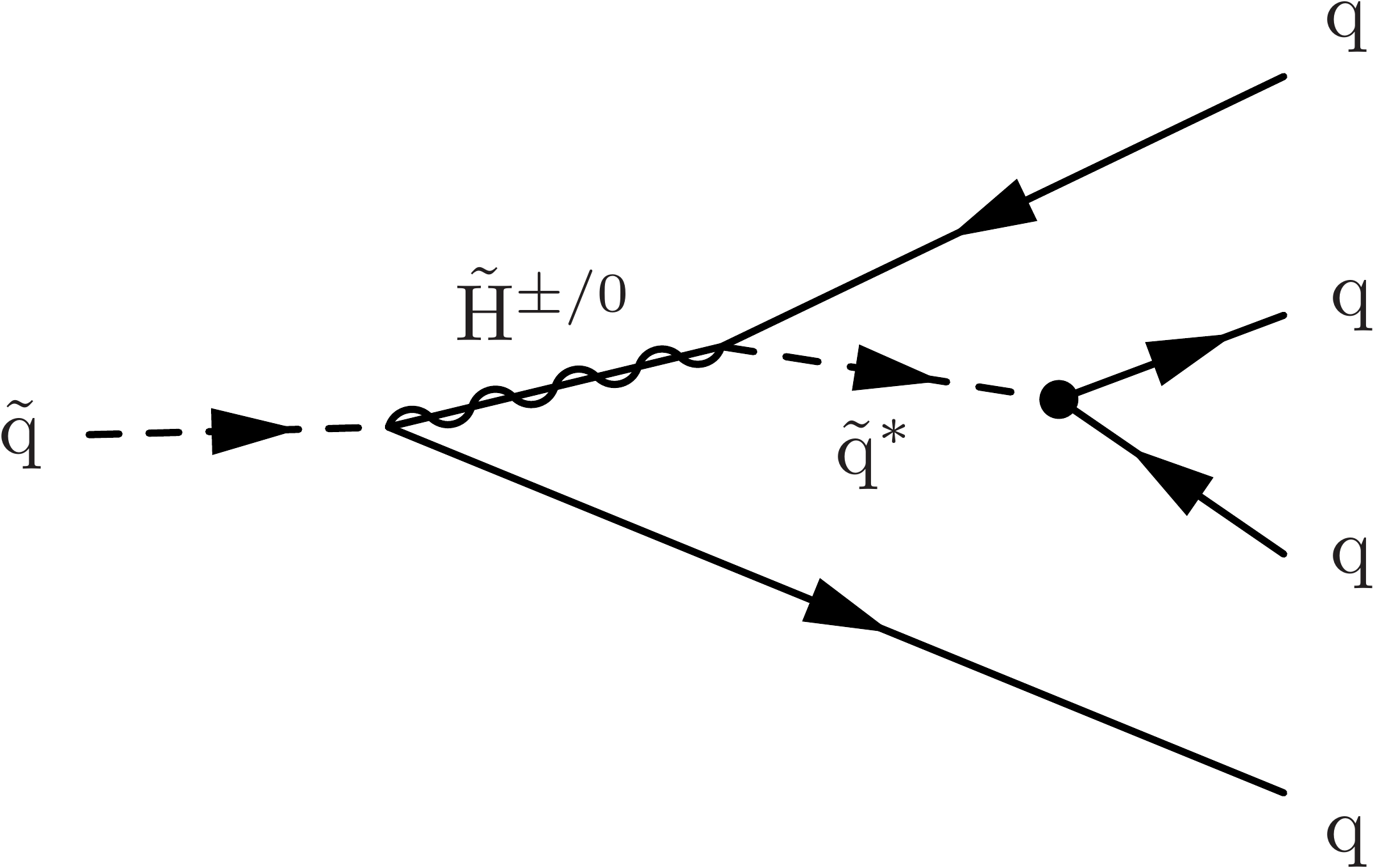}
	\hspace{1.5mm}
	\includegraphics[width=0.95\cmsFigWidth]{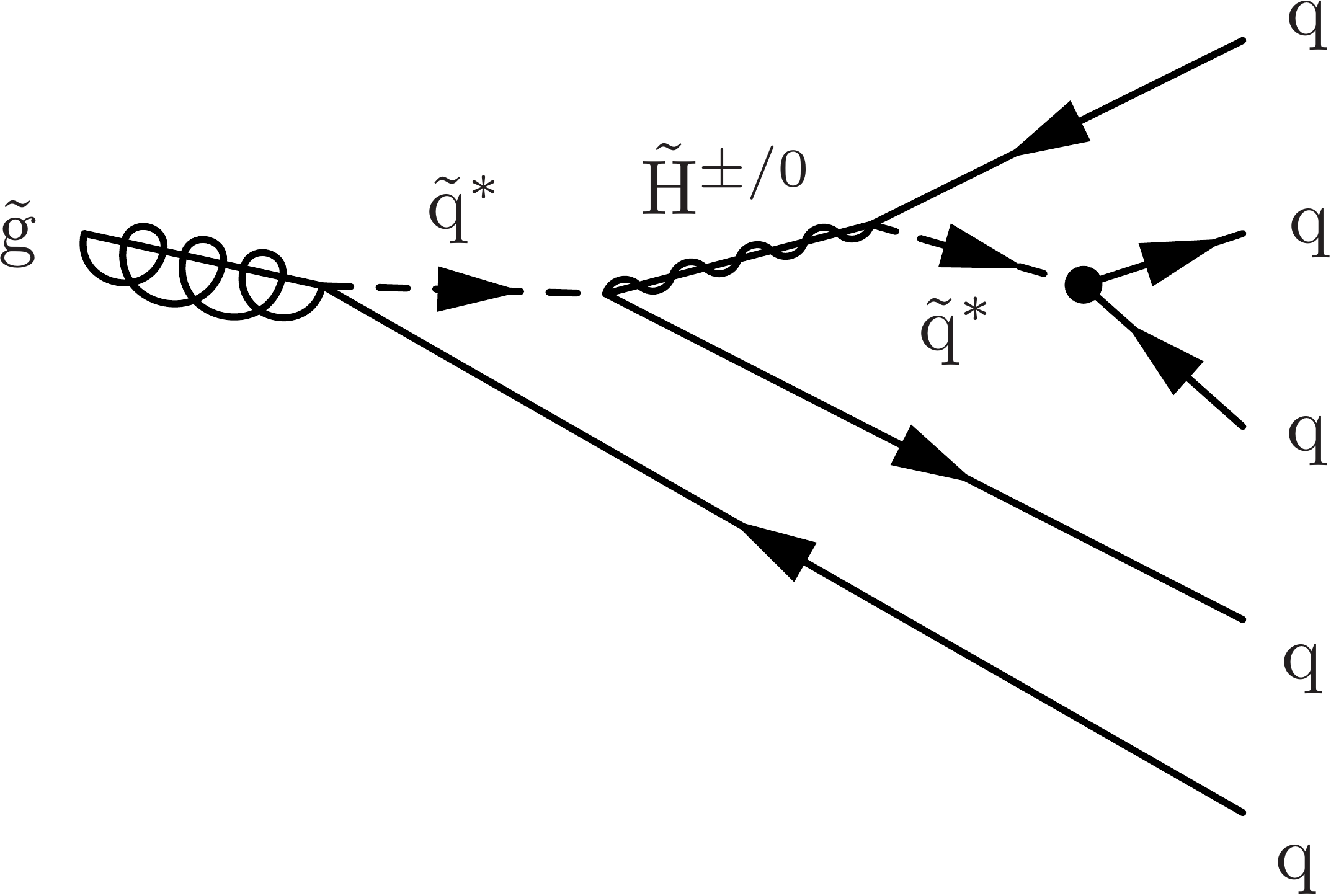}
	\caption{Leading order Feynman diagrams for a squark decaying into four quarks via an intermediate higgsino and a three-body decay involving an off-shell squark with an RPV coupling (left) and for a gluino decaying into five quarks via a three-body decay into two quarks and a higgsino, which decays into three quarks as in the squark case (right). This analysis assumes that the initial squarks (or gluinos) are pair-produced and that they each decay as presented.}
	\label{fig:feyn}
\end{figure}

This analysis probes a previously unexamined region of the SUSY model space, where pair-produced squarks with masses as light as 100\GeV each decay into four non-top quarks through hadronic RPV couplings~\cite{BARBIER20051}.  Such processes proceed through a two-body decay of the squark into a higgsino and a quark, and the higgsino subsequently undergoes an RPV three-body decay via an off-shell squark (\squarkoff) into quarks, as shown in Fig.~\ref{fig:feyn} (left). This decay mode is expected to dominate over direct squark decays into two quarks if the higgsino mass is less than the squark mass (as expected in natural scenarios) and if the RPV coupling is small. A related topology with five-quark resonances occurs when the gluino undergoes a three-body decay into two quarks and a higgsino, and the higgsino decays into three quarks, as shown in Fig.~\ref{fig:feyn} (right). Searches by the experiments at LEP strongly disfavor squark and (when degenerate with the chargino) higgsino masses below 100\GeV, while existing LHC searches are only sensitive to multijet scenarios above at least 400\GeV~\cite{cms-exo-13-001,cms-sus-14-003,atlas-susy-160219,atlas-susy-150219}. Light pair-produced squarks and gluinos that decay through RPV couplings into two and three quarks, respectively, have also been directly constrained by Tevatron and LHC searches~\cite{cms-exo-17-001,Aaboud:2017nmi,cms-exo-12-052,atlas-susy-160127}.

We take advantage of the large cross section predicted for squark pair production, focusing on the small fraction of phase space in which the squarks are produced back-to-back and the decay products of each squark reside entirely within a single large jet. This requirement reduces the combinatorial background and recovers the resonant structure of the squark decay. Furthermore, the masses of the two jets are required to be approximately equal, and the jets must evince substructure consistent with at least four non-top quarks. These conditions help to suppress the immense background of light-quark jets produced by the strong interaction, referred to as quantum chromodynamics (QCD) multijet events, which is estimated from data with a novel technique. The background from top quark-antiquark (\ttbar) events is also significant; its contribution is estimated from simulation. The \ttbar events in control regions are used in the estimation of the selection efficiency, jet mass scale, and jet mass resolution.

The motivation for this search comes from natural SUSY, but the analysis aims more broadly for new phenomena in general. We make no assumptions about the helicity, the intermediate decay topology, or the presence of secondary vertices (characteristic of heavy-flavor quarks) in the final state, so as to be more inclusive and to focus on the more experimentally challenging light-quark mode. The reconstruction of the final state treats charged leptons and photons on equal footing with quarks and gluons, resulting in an analysis that is sensitive to a broad class of phenomena.  Many solutions to the hierarchy problem invoke new, sub-TeV colored particles, and this analysis, using a novel background estimation technique, can provide strong constraints if these particles are to decay primarily into large numbers of detectable objects.

The CMS detector consists of a silicon tracker, a lead-tungstate electromagnetic calorimeter, an interleaved brass and plastic scintillator hadron calorimeter, and gas-ionization muon detectors. A superconducting solenoid provides a uniform magnetic field within the detector. Potentially interesting observed events are selected by a two-tiered trigger system~\cite{trigger}. A more detailed description of the CMS detector can be found in Ref.~\cite{cms_2008}.

Observations from each of the CMS detector components are reconstructed and combined by particle flow (PF) algorithms~\cite{pf} into particle objects. These objects are clustered into jets using the \FASTJET software package~\cite{fastjet} with either the anti-\kt (AK) or Cambridge--Aachen (CA) algorithm with a size parameter $R$~\cite{jetography}. A collection of jets is referenced by the acronym for the clustering algorithm used, followed by ten times the associated value of $R$: for example, ``CA12'' refers to jets clustered with the CA algorithm with $R=1.2$, the jet collection primarily used in this analysis.

We analyze data collected by the CMS detector in 2015-2016 from proton-proton collisions at ${\sqrt{s}=13\TeV}$ and corresponding to an integrated luminosity of 38.2\fbinv. Events considered for analysis were selected by a trigger that imposes a requirement on the event ${\HT\equiv\sum_i p_{\mathrm{T}, i}}$, where $i$ iterates over the AK8 jets in the event with transverse momentum ${\pt>150\GeV}$ and ${\abs{\eta}<2.5}$, where $\eta$ is the pseudorapidity.

{\tolerance=8000
Signal events are generated at leading order in perturbative QCD using \MGvATNLO 2.5.5~\cite{madgraph5} with the NNPDF2.3 parton distribution function~\cite{nnpdf} and \PYTHIA~8.212~\cite{pythia82} with the underlying event tune CUETP8M1~\cite{pythia:tune} to simulate the subsequent parton showering and decays. No additional partons are consider in the initial state. Events for 11 squark mass points between 0.1 and 0.8\TeV are produced. In each case, the higgsino mass is set to 75\% of the squark mass in order to distribute the squark energy evenly among the daughter quarks, and the gluino is assumed to be too massive to participate in the process. Simulated gluino signal events are produced for 17 gluino mass points between 0.1 and 1.5\TeV. In this case, the higgsino mass is set to 60\% of the gluino mass, and the squarks are assumed to be decoupled. Squark and gluino events are normalized to the theoretical cross sections calculated up to next-to-leading-order plus next-to-leading-logarithm precision~\cite{susyxs}. The squark cross sections are chosen to be those computed for top squarks.
\par}

Background from \ttbar production is generated with \POWHEG~v2~\cite{powheg_2004,powheg_2007,powheg_2010}, and the \pt distribution of the top quarks is reweighted to agree with measurement~\cite{ptreweight}. Background from QCD multijet events is simulated at leading order using \MGvATNLO~2.2.2 with up to four partons included in the matrix-element calculation. Backgrounds such as hadronic \PW{+}jets and \PZ{+}jets make a negligible contribution and are absorbed into the QCD multijet background estimate. All background samples are generated with NNPDF2.3 or NNPDF3.0 parton distribution functions and are interfaced with \PYTHIA for the computation of fragmentation and hadronization. Additional proton-proton interactions within the same or a nearby bunch crossing (pileup) are added to the simulated events, with a frequency distribution chosen to match that observed in data.

In order to capture as many of the final-state constituents of the squark decay products as possible in a single jet, we cluster the PF particle objects of each event into CA12 jets. To reduce the effect of pileup interactions, we ignore any charged hadrons not associated with the primary vertex in this clustering procedure. The primary vertex is chosen to be the vertex with the largest quadratic sum of the \pt values of AK4 jets clustered from tracks assigned to the vertex and the \pt corresponding to the negative vector-\pt sum of these AK4 jets. The CA12 jet energies are corrected to compensate for the combined response function of the CMS calorimeters and the presence of neutral hadrons from pileup, with calibrations designed for AK8 jets~\cite{JEC}. The data are corrected with an additional scale factor to account for the residual difference in the detector response between simulation and data.
Miscorrections that result from the application of AK8 calibrations to the CA12 jets are measured and are accounted for in the final results, together with their uncertainties.
Finally, to reject misreconstructed CA12 jets or CA12 jets derived from calorimeter noise~\cite{hcalnoise}, we impose restrictions on the neutral and charged CA12 jet constituents by applying loose jet identification criteria~\cite{jetid}.

We analyze the substructure of the CA12 jets using the $N$-subjettiness variables, denoted $\tau_N$~\cite{nsub}. These quantities are determined by reclustering the constituents of the jet with the \kt algorithm~\cite{jetography} into $N$ subjets and then calculating
\begin{linenomath}
	\begin{align}\label{eq:tau}
		\tauN\equiv\frac{1}{d_0}\sum_i p_{\mathrm{T},i}\min{\left\{\Delta R_{1,i},\,\dots,\,\Delta R_{N,i}\right\}},
	\end{align}
\end{linenomath}
where $i$ iterates over the jet constituents, ${d_0\equiv\sum_i p_{\mathrm{T},i}R}$ with $R$ representing the size parameter of the jet, and $\Delta R_{j,i}$ is the distance in $\eta$-$\phi$ space between the $j^\mathrm{th}$ subjet and the $i^\mathrm{th}$ jet constituent, where $\phi$ is the azimuthal angle. The lower the value of \tauN for a jet, the more the jet is consistent with containing $N$ partons. The relative $N$-subjettiness $\taukl\equiv\tau_k/\tau_l$, for specific integers $k$ and $l$, characterizes the substructure of a jet. In particular, \taufourtwo is used to discriminate against QCD multijet events and \taufourthree to discriminate against \ttbar events.

Finally, we apply a pruning algorithm~\cite{pruning} that reclusters the original constituents of a jet using a CA algorithm modified to ignore any objects with small relative \pt ($z_\mathrm{cut} = 0.1$) and large kinematically weighted displacement ($D_\mathrm{cut} = 0.5$). This refinement serves to suppress perturbative radiation characteristic of QCD multijet events as well as to reduce the contribution to jet masses from detector noise, pileup, and the underlying event. In this Letter, the term ``fat jet'' is used to refer to a CA12 jet with $\tauN$ requirements, and the mass of a jet $m$ always refers to the mass after pruning.

We consider events with $\HT>0.9\TeV$ in order to guarantee that we analyze a kinematic region with a fully efficient trigger. In addition, we require that the two \pt-leading CA12 jets of the events must each satisfy the following three conditions: $\pt>0.4\TeV$ to ensure that the jets are more likely to contain all signal products, $\abs{\eta}<2$ to require that the jets are dominated by objects in the central detector region, and $\tautwoone<0.75$ to select CA12 jets with potential for substructure.

From these selected events, we define a signal region (SR) and two control regions (CRs). The regions are formed by selecting events in which the two \pt-leading CA12 jets (indexed 1 and~2) each satisfy certain requirements on \taufourtwo and \taufourthree, called the region's ``fat-jet tag,'' and also pass relational criteria of ${\masy\equiv\abs{m_1-m_2}/\left(m_1+m_2\right)}<0.1$ and ${\deta\equiv\abs{\eta_1-\eta_2}}<1.0$, defining a ``fat-jet pair.'' The SR fat-jet tag requires ${\taufourtwo<0.50}$ and ${\taufourthree<0.80}$, choices that were optimized for the squark signal significance. We define an inclusive CR and a \PQb-tagged CR, which we use to verify the closure of the background prediction in data and to estimate the systematic uncertainties that are further constrained in the SR. Both CR fat-jet tags impose the same $N$-subjettiness requirements on each fat jet: ${\taufourtwo<0.55}$, ${\taufourthree<0.90}$, and either ${\taufourtwo>0.50}$ or ${\taufourthree>0.80}$. The \PQb-tagged CR fat-jet tag additionally requires a loose \PQb tag~\cite{btag} (with approximately 80\% tagging efficiency and 9\% mistag rate) in order to select a \ttbar-enriched sample. The CA12 jets are \PQb tagged by matching them to the \PQb-tagged AK4 jets that are closest in $\eta$-$\phi$ space, assigning to each CA12 jet the smallest \PQb-tagging discriminator from among the matched AK4 jets. The final discriminating variable of this analysis is the average pruned mass of the two \pt-leading fat jets in an event: ${\mavg\equiv\left(m_1+m_2\right)/2}$.

The fraction of squark events in the SR ranges from $<0.001\%$ for the 0.1\TeV squark mass point to 1\% for the 0.8\TeV mass point. For very light squarks, the small selection rate is predominantly driven by the low kinematic acceptance, but it is compensated for by a large production cross section (1520\unit{pb} for 0.1\TeV top squarks). Similar values for the selection efficiency are found for gluino events, although the \mavg distribution for gluinos is broader than for squarks in part because of the extra quark in the decay chain. These properties are illustrated in \suppMaterial: Fig.~\suppMaterialRef{1}{\ref{fig:accept}} shows the fraction of signal events in the SR, and Figs.~\suppMaterialRef{2}{\ref{fig:signals_sq}} and \suppMaterialRef{3}{\ref{fig:signals_sg}} show \mavg distributions of select squark and gluino mass points, respectively.

We estimate the contribution of a particular signal to the observed events by performing a maximum-likelihood fit of the combined \mavg probability density functions (PDFs) of the signal and the two backgrounds to the \mavg distribution of the data.
The signal and \ttbar background \mavg PDFs are taken from simulation; the QCD multijet background \mavg PDF is constructed using a data-driven method detailed below. Each \mavg PDF is assigned three fit parameters: a normalization factor, a shift of the median of the PDF from the nominal value, and a width factor (stretch) about the median of the PDF. These fit parameters allow us to account for observed shape differences between simulation and data.

\begin{figure}[htbp]
	\centering
	\includegraphics[width=\cmsFigWidth]{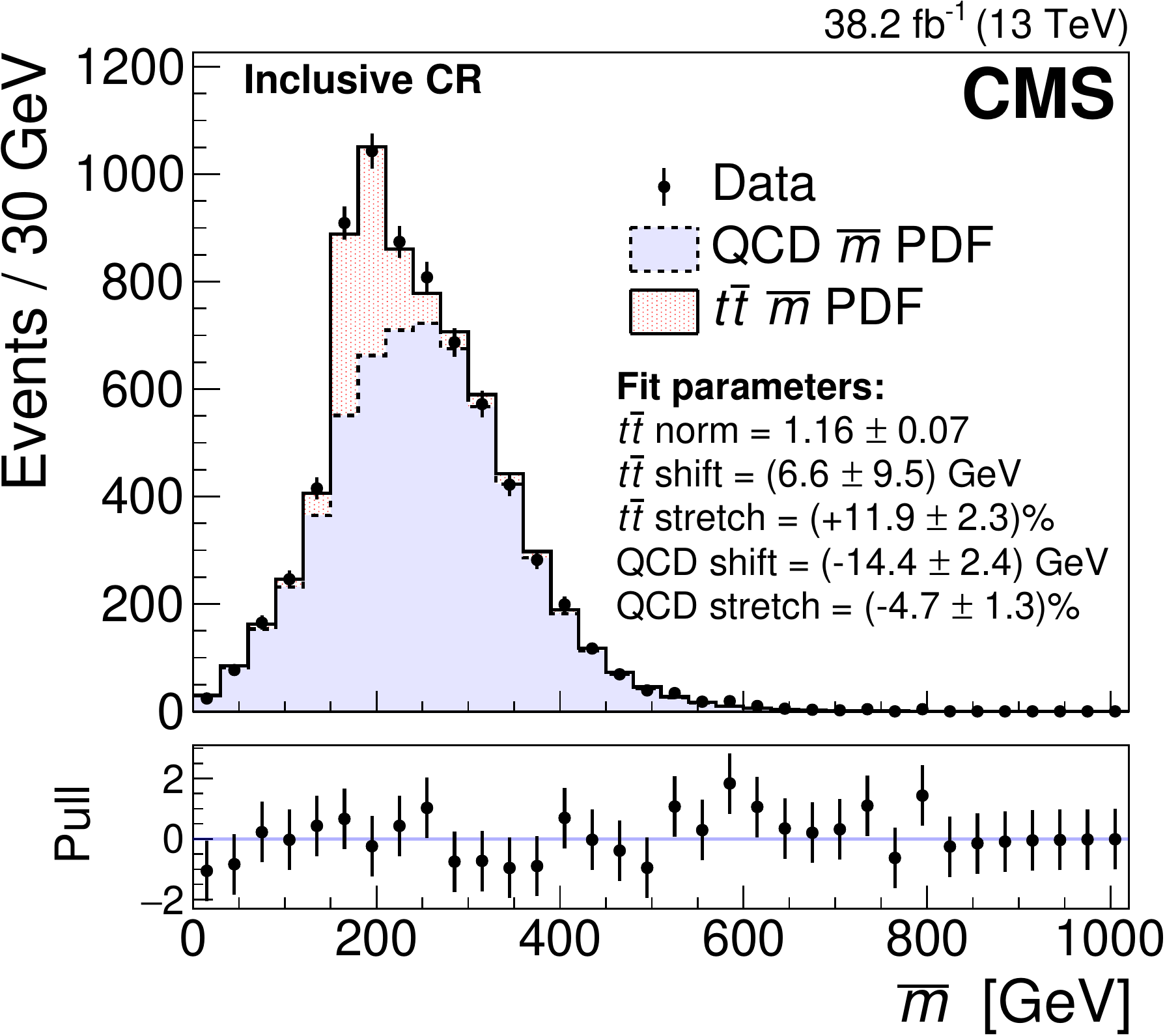}
	\includegraphics[width=\cmsFigWidth]{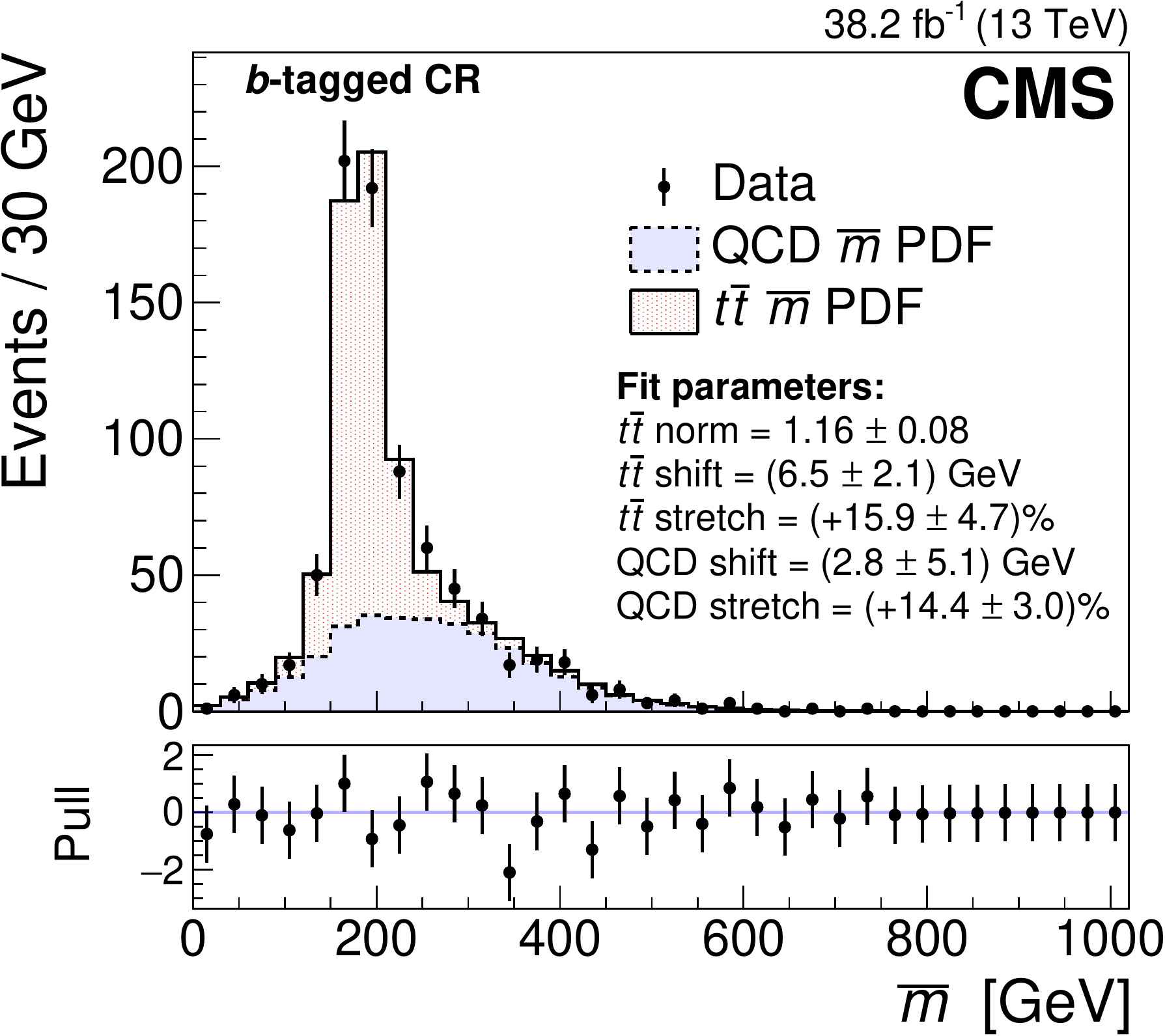}
	\caption{The background estimation results in the inclusive CR (left) and \PQb-tagged CR (right). The relevant fit parameters are displayed on each plot. The pull of each \mavg bin is the difference of the value of the data \mavg distribution and the value of the combined background post-fit \mavg PDF, divided by the statistical uncertainty.}
	\label{fig:closure_data}
\end{figure}

This procedure is applied to the CRs in the data, and we use the post-fit values of the \mavg PDF parameters as estimates of the background systematic uncertainties. Figure~\ref{fig:closure_data} shows the post-fit background \mavg PDFs and the associated fit parameters, assuming no signal contribution. Signal contamination in the CRs is generally small, and we repeat the same procedure including the \mavg PDF for each signal mass point to establish that the presented null hypothesis results in the best fit.
The \ttbar \mavg PDF fit parameters have physical interpretations: the normalization corresponds to the data-to-simulation fat-jet tagging scale factor, the shift corresponds to the tagged fat-jet mass scale, and the stretch corresponds to the tagged fat-jet mass resolution. As presented in Fig.~\ref{fig:closure_data}, the shifts in the CRs are generally consistent with zero, but the stretches imply broadening, consistent with the known ${\sim}10\%$ underestimation of the jet energy resolution in simulation~\cite{JEC}. We study a variety of alternative choices of CR fat-jet tags, which similarly demonstrate good closure.
The background estimation method reproduces the data in the inclusive CR, which has greater statistical precision than the SR, lending confidence that it provides reliable results in the SR as well.

We derive the QCD multijet background \mavg PDF from a large sample of data events in order to avoid the mismodeling of simulated QCD multijet events. For this purpose, events are selected in which at least the \pt-leading jet satisfies the fat-jet tag; the number of such events is more than an order of magnitude greater than the number required to have a fat-jet pair. We form a collection of fat jets by taking the \pt-leading fat jet from each of these events. From this collection of fat jets from different events, we form all possible fat-jet pairs. The \mavg distribution of these artificial fat-jet pairs provides an \mavg PDF for QCD multijet events. This construction treats the $m$ distribution of \pt-leading fat jets as an $m$ PDF, denoted \pdfm, from which two fat-jet masses are sampled to form an \mavg PDF \pdfmavg:
\begin{linenomath}
	\begin{equation}\label{eq:P} P_{\text{avg}}\hspace{-0.3mm}(\mavg)\,{=}\!\int_0^{2\mavg}\hspace{-3mm}P(x)P(2\mavg-x)\,\theta\hspace{-0.8mm}\left(\hspace{-0.5mm}0.1-\left|\frac{x-\mavg}{\mavg}\right|\right)\hspace{-0.8mm}\rd{}x,
	\end{equation}
\end{linenomath}
where $\theta$ is the Heaviside step function that imposes the \masy fat-jet pair requirement. Although the \deta fat-jet pair requirement does not appear explicitly in Eq.~\eqref{eq:P}, it is imposed when we implement this construction.

Fat jets from QCD multijet events do not contain the products of partons with well-defined masses, and so the \pdfm derived from them has a significant \pt dependence that must be taken into account. We accommodate this effect by constructing a different \pdfmavg for consecutive \HT windows, which we then combine with weights chosen to reproduce to the observed \HT distribution in data.
Figure~\suppMaterialRef{4}{\ref{fig:technique}} of \suppMaterial shows \pdfm and the derived \pdfmavg with and without this \HT reweighting, demonstrating the significant effect of the reweighting on the high-\mavg tail of the distribution.
Corrections for the contamination of the QCD multijet \mavg PDF by signal or other backgrounds are unnecessary, given the dominance of QCD multijet events over all other processes in the events used to construct \pdfm.

\begin{figure}[htbp]
	\centering
	\includegraphics[width=\cmsFigWidth]{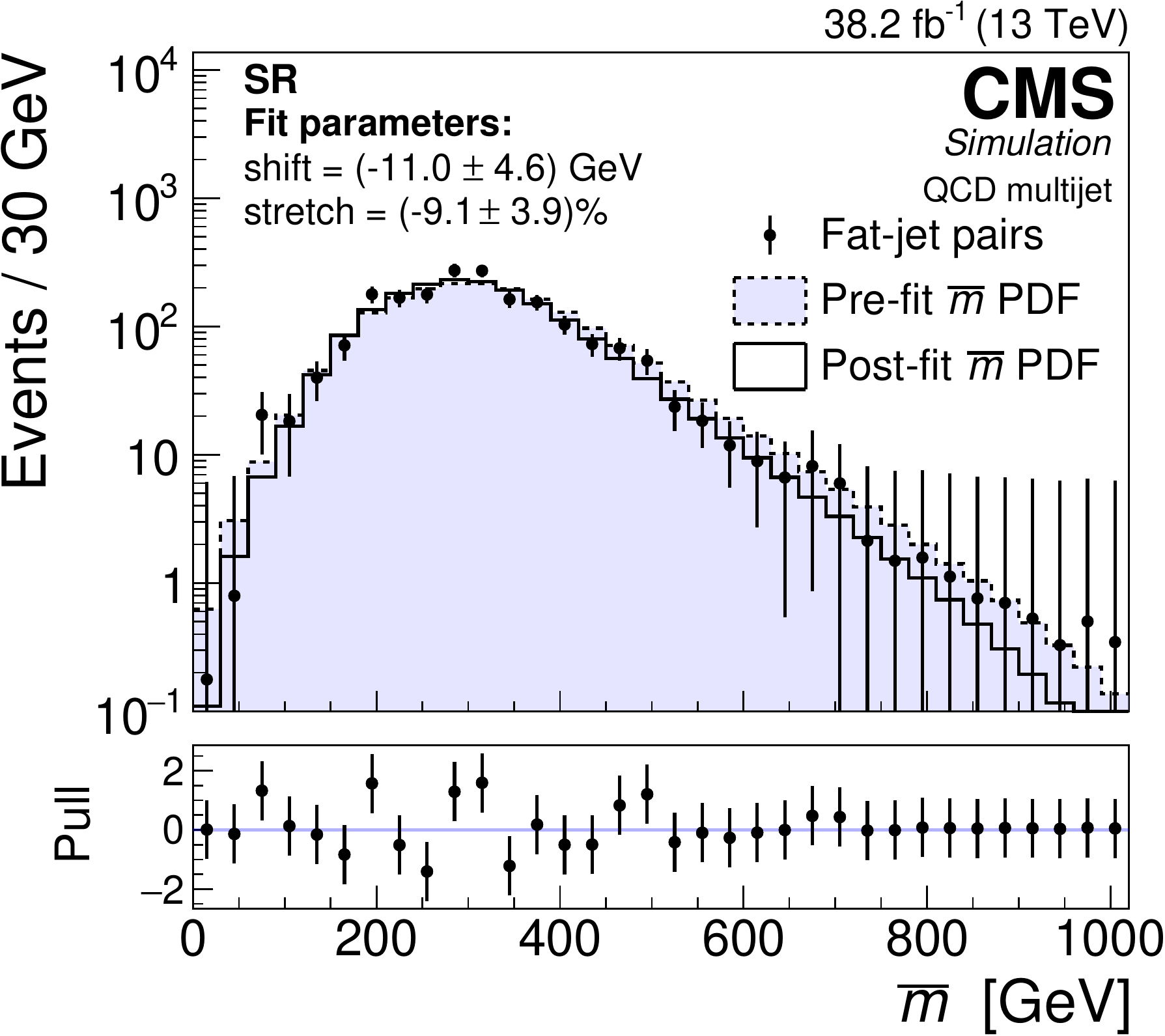}
	\includegraphics[width=\cmsFigWidth]{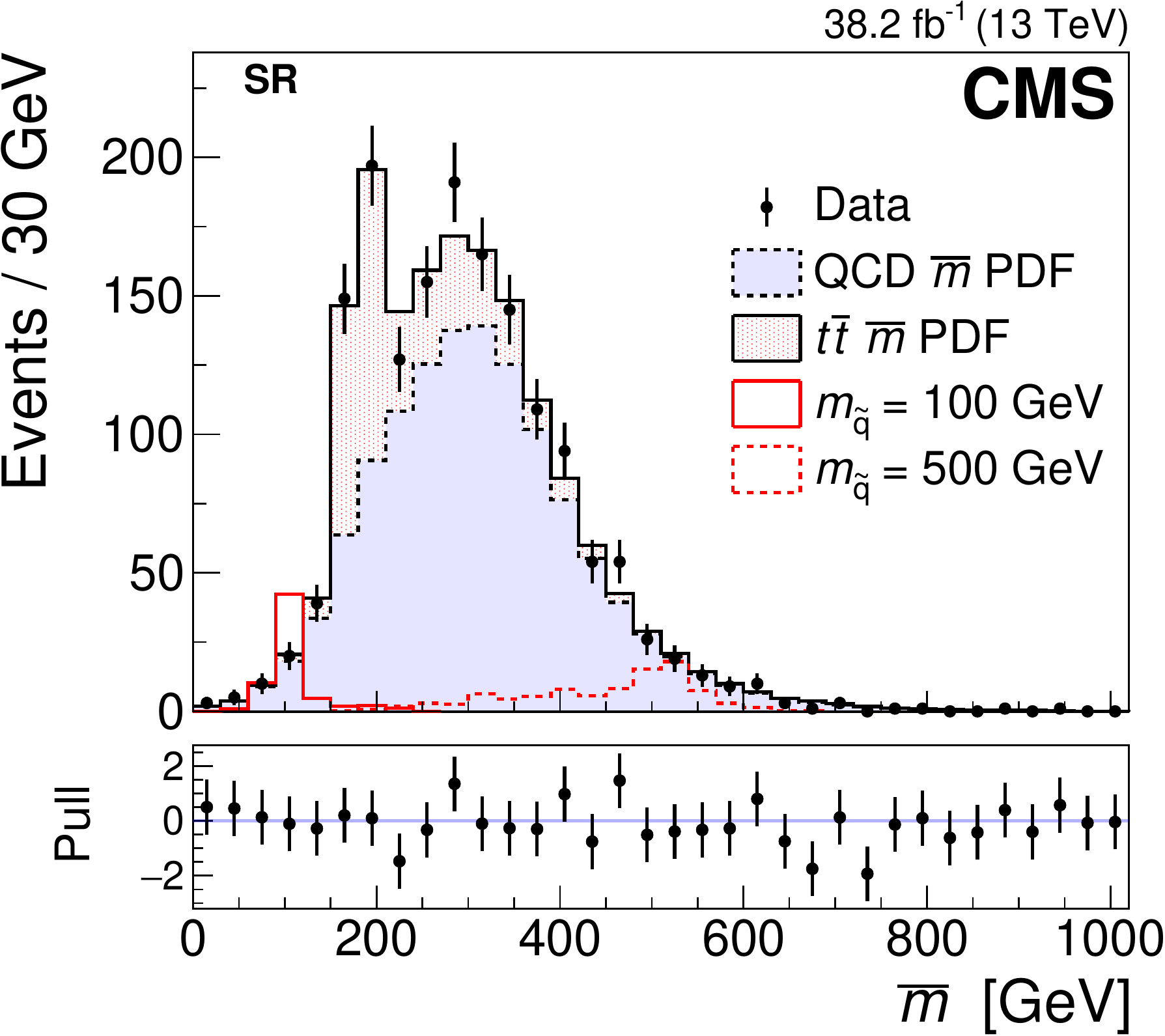}
	\caption{Left: a comparison between the QCD multijet \mavg PDF and the tagged fat-jet pair selection in simulated QCD multijet events that pass the signal selection. The pull of each \mavg bin is the difference of the value of the fat-jet pair \mavg distribution and the value of the post-fit \mavg PDF, divided by the statistical uncertainty. Right: distributions in \mavg, including the predicted (post-fit) background contributions. Shown also are the signals from squarks with masses of 100\GeV and 500\GeV. The pull of each \mavg bin is the difference of the value of the data \mavg distribution and the value of the background prediction, divided by the statistical uncertainty.}
	\label{fig:sr}
\end{figure}

The accuracy of this QCD multijet \mavg PDF construction is demonstrated by a closure test. We construct QCD multijet \mavg PDFs from simulated QCD multijet events in the SR and CRs and compare them to the \mavg distributions of the same regions' tagged fat-jet pairs (as reconstructed by the simulation); the result for the SR is shown in Fig.~\ref{fig:sr} (left). The SR \mavg PDF only needs to be shifted downward by approximately 11\GeV and compressed by about 9\% to match the simulated fat-jet pair distribution. Similar tests with alternative choices for the $N$-subjettiness variables and different simulation procedures for the QCD multijet events also demonstrate good closure.

Using the data in the SR, we apply a Bayesian construction with a flat signal-strength prior~\cite{CowanPDGStat} to compute 95\% confidence level (\CL) upper limits on squark (gluino) pair production and decay to four (five) quarks. Markov chain Monte Carlo techniques~\cite{MCMC} are used to integrate out the nuisance parameters. We analyze the data in 30\GeV bins, as shown in Fig.~\ref{fig:sr} (right).

We assign nuisance parameters corresponding to the shift, stretch, and normalization of the QCD multijet, \ttbar, and signal \mavg PDFs. The nuisance parameters have Gaussian PDFs. The QCD multijet \mavg PDF shape nuisance parameter uncertainties are taken from the CRs, but no initial shift or stretch is applied, since these parameters in the SR might be different from those in the CRs. We do not have a robust prediction for the total rate of QCD multijet events in the SR, so the QCD multijet \mavg PDF normalization is chosen to be a free parameter. This procedure is reasonable because the QCD multijet shape is much broader than that of the signal, allowing the data in the SR to directly constrain this background.

The \ttbar rate and shape nuisance parameter uncertainties are determined by their measured values in the CRs; we start the minimization procedure with the initial value of these parameters reflecting no shift or stretch, as that is consistent with measurement. Interpreting the \ttbar shift and stretch as jet-mass scale and resolution uncertainties, respectively, we apply the same uncertainties to the signal shape by constraining them to match the \ttbar background nuisance parameters. For the signal rate, we assign the same uncertainty value that we measured for \ttbar events, since the signal and \ttbar production and decay topologies are similar. The uncertainty in the signal rate includes contributions from the integrated luminosity and the fat-jet tagging scale factor. Uncertainties in the signal acceptance due to the parton distribution functions of the proton contribute subdominantly. The central values and standard deviations of all of the nuisance parameters are shown in Table \ref{table:nuisance}.

\begin{table}[htbp]
	\centering
	\topcaption{The nuisance parameters corresponding to each rate and shape parameter of the background and signal distributions, before and after the maximum-likelihood fit. Except for the QCD multijet \mavg PDF normalization, which is floating (and the value of which is simply the event yield with its statistical uncertainty), each nuisance parameter has a Gaussian prior PDF and is reported with the given mean and standard deviation.}
	\begin{scotch}{lr@{}lr@{}l}
		{Parameter} & \multicolumn{2}{c}{{Pre-fit value}} & \multicolumn{2}{c}{{Post-fit value}}\\
		\hline
		QCD normalization [events] & \multicolumn{2}{c}{\hspace{-0.0 in}floating} & $1222$ & $\:\pm\: 35$\\
		QCD shift [\GeVns{}] & $0$ & $\:\pm\: 17$ & $-8$ & $\:\pm\: 4$\\
		QCD stretch & $(0$ & $\:\pm\: 18)\%$ & $(-1$ & $\:\pm\: 3)\%$\\
		\ttbar normalization & $1.00$ & $\:\pm\: 0.24$ & $1.08$ & $\:\pm\: 0.14$ \\
		\ttbar and signal shift [\GeVns{}] & $0$ & $\:\pm\: 16$ & $-10$ & $\:\pm\: 6$\\
		\ttbar and signal stretch & $(0$ & $\:\pm\: 20)\%$ & $(15$ & $\:\pm\: 9)\%$\\
		Signal normalization & $1.00$ & $\:\pm\: 0.24$ & \multicolumn{2}{c}{\hspace{-0.15 in}\NA}\\
	\end{scotch}
	\label{table:nuisance}
\end{table}

\begin{figure}[htbp]
	\centering
	\includegraphics[width=\cmsFigWidth]{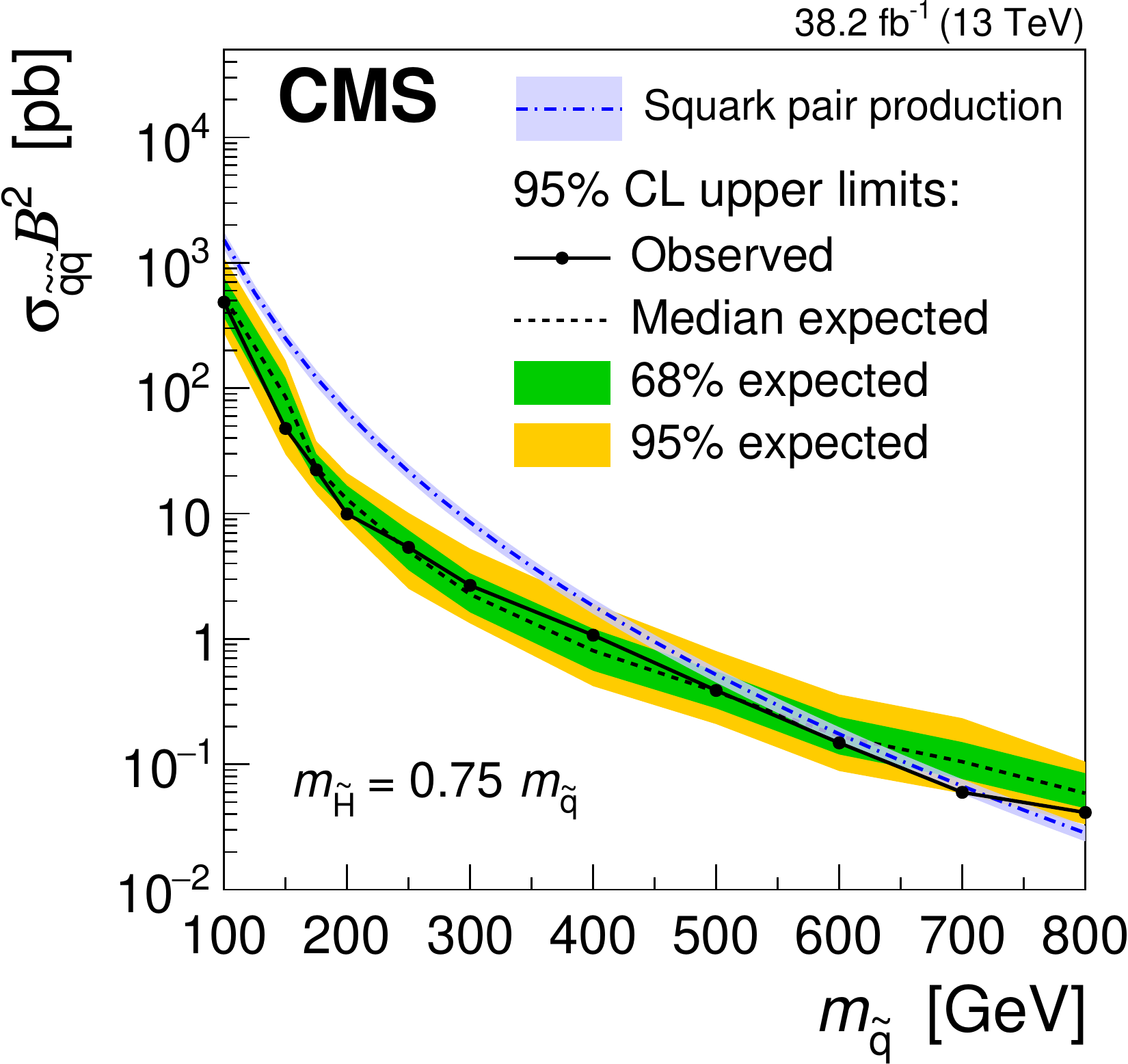}
	\includegraphics[width=\cmsFigWidth]{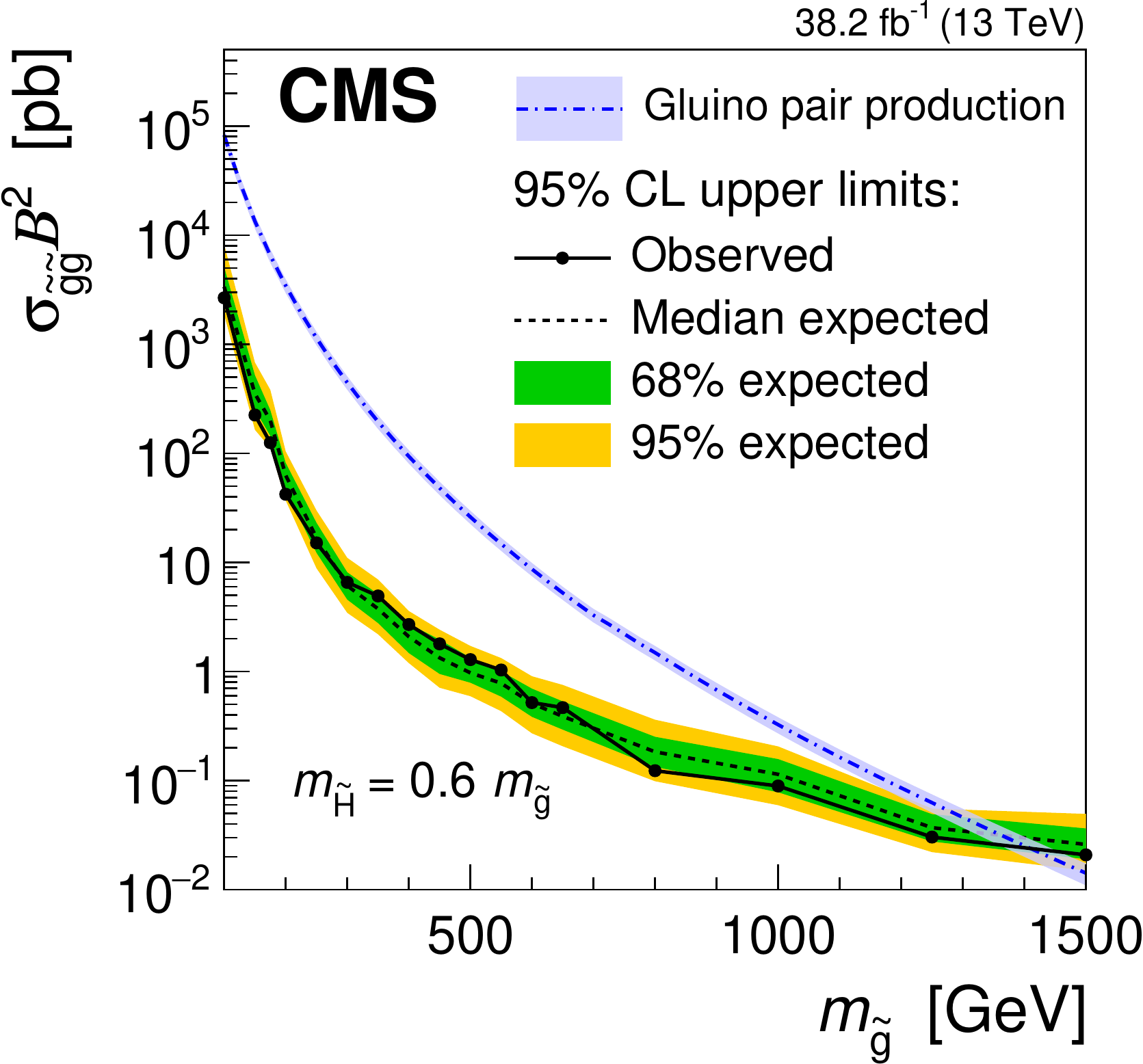}
	\caption{The expected and observed limits on the product of the pair-production cross section and branching fraction squared for a quark decaying to four quarks (left) and for a gluino decaying to five quarks (right).}
	\label{fig:results}
\end{figure}

The resulting limits on the production cross section and the predicted background components in the SR are shown in Fig.~\ref{fig:results}. Assuming the top squark production cross section, squark masses between 0.10 and 0.72\TeV are excluded at 95\% \CLndot. Gluinos decaying to five quarks with masses between 0.10 and 1.41\TeV are also excluded at 95\% \CLndot. The post-fit total background estimate agrees with the data. The posterior distributions of the nuisance parameters confirm that the background component predictions are not significantly different from the estimates.

In summary, a search has been conducted at the LHC for light pair-produced resonances that each decay into at least four quarks. No statistically significant excess over the expectation is observed. The data impose limits on $R$-parity-violating supersymmetry pair production~\cite{Evans2013}, excluding squark masses between 0.10 and 0.72\TeV and gluino masses between 0.10 and 1.41\TeV. These are the first constraints that have been placed on pair-produced particles with masses below 400\GeV that decay into four or five quarks, bridging a significant gap in the coverage of $R$-parity-violating supersymmetry parameter space.
This analysis is sufficiently general that it can be applied to other models describing pair-produced particles decaying into four or more detectable objects.

\begin{acknowledgments}
We congratulate our colleagues in the CERN accelerator departments for the excellent performance of the LHC and thank the technical and administrative staffs at CERN and at other CMS institutes for their contributions to the success of the CMS effort. In addition, we gratefully acknowledge the computing centers and personnel of the Worldwide LHC Computing Grid for delivering so effectively the computing infrastructure essential to our analyses. Finally, we acknowledge the enduring support for the construction and operation of the LHC and the CMS detector provided by the following funding agencies: BMBWF and FWF (Austria); FNRS and FWO (Belgium); CNPq, CAPES, FAPERJ, FAPERGS, and FAPESP (Brazil); MES (Bulgaria); CERN; CAS, MoST, and NSFC (China); COLCIENCIAS (Colombia); MSES and CSF (Croatia); RPF (Cyprus); SENESCYT (Ecuador); MoER, ERC IUT, and ERDF (Estonia); Academy of Finland, MEC, and HIP (Finland); CEA and CNRS/IN2P3 (France); BMBF, DFG, and HGF (Germany); GSRT (Greece); NKFIA (Hungary); DAE and DST (India); IPM (Iran); SFI (Ireland); INFN (Italy); MSIP and NRF (Republic of Korea); MES (Latvia); LAS (Lithuania); MOE and UM (Malaysia); BUAP, CINVESTAV, CONACYT, LNS, SEP, and UASLP-FAI (Mexico); MOS (Montenegro); MBIE (New Zealand); PAEC (Pakistan); MSHE and NSC (Poland); FCT (Portugal); JINR (Dubna); MON, RosAtom, RAS, RFBR, and NRC KI (Russia); MESTD (Serbia); SEIDI, CPAN, PCTI, and FEDER (Spain); MOSTR (Sri Lanka); Swiss Funding Agencies (Switzerland); MST (Taipei); ThEPCenter, IPST, STAR, and NSTDA (Thailand); TUBITAK and TAEK (Turkey); NASU and SFFR (Ukraine); STFC (United Kingdom); DOE and NSF (USA).
\end{acknowledgments}

\bibliography{auto_generated}

\ifthenelse{\boolean{cms@external}}{}{
	\clearpage
	\appendix
	\numberwithin{figure}{section}
	\section{Figures illustrating properties of simulated signal events and background estimation\label{app:suppMat}}
	
\begin{figure}[htb!]
	\centering		\includegraphics[width=0.48\textwidth]{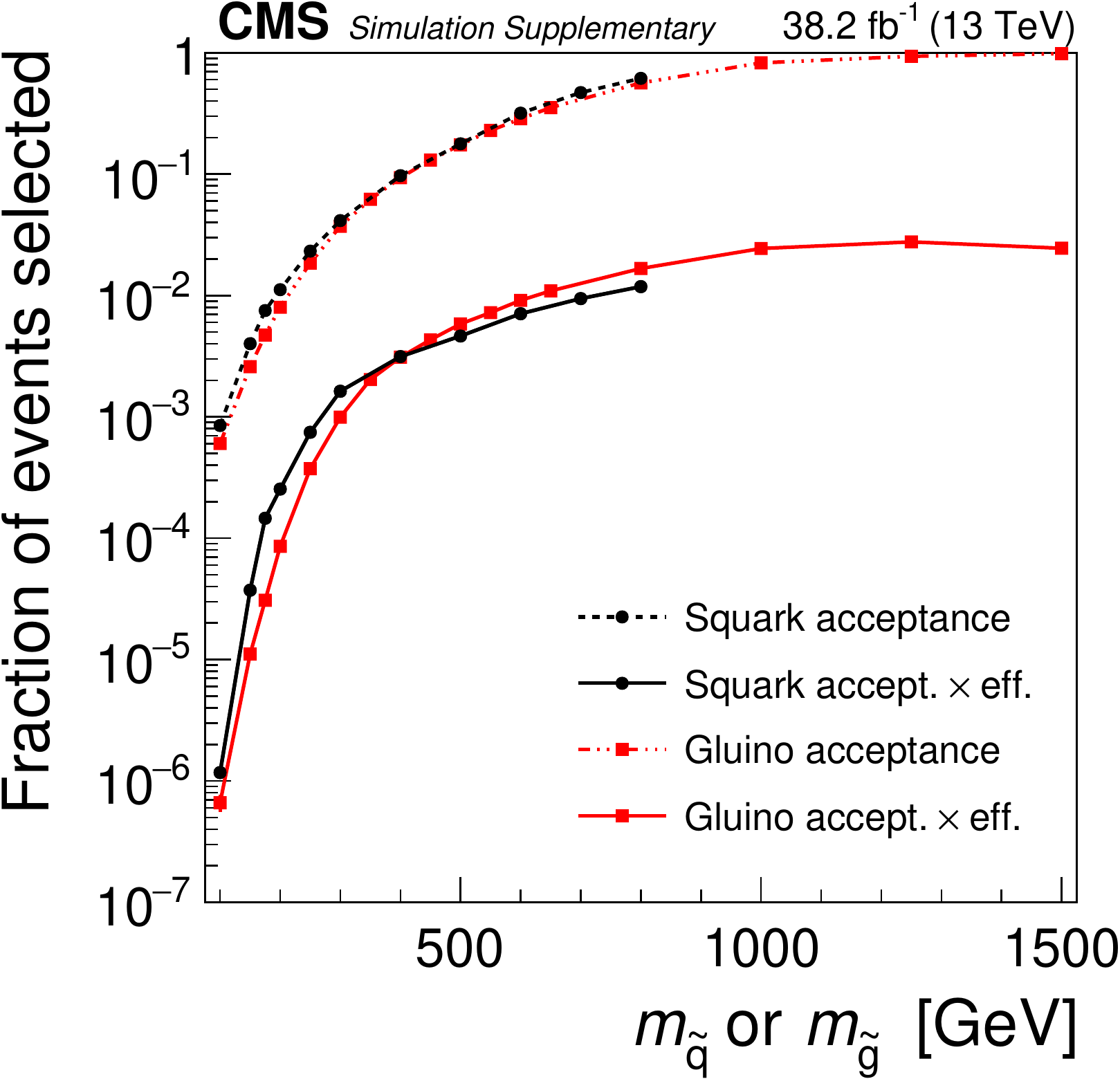}
		\caption{Acceptance and acceptance times efficiency in the SR for the simulated squark and gluino signal samples.}
		\label{fig:accept}	
\end{figure}

\begin{figure}[htb!]
	\centering
		\includegraphics[width=0.48\textwidth]{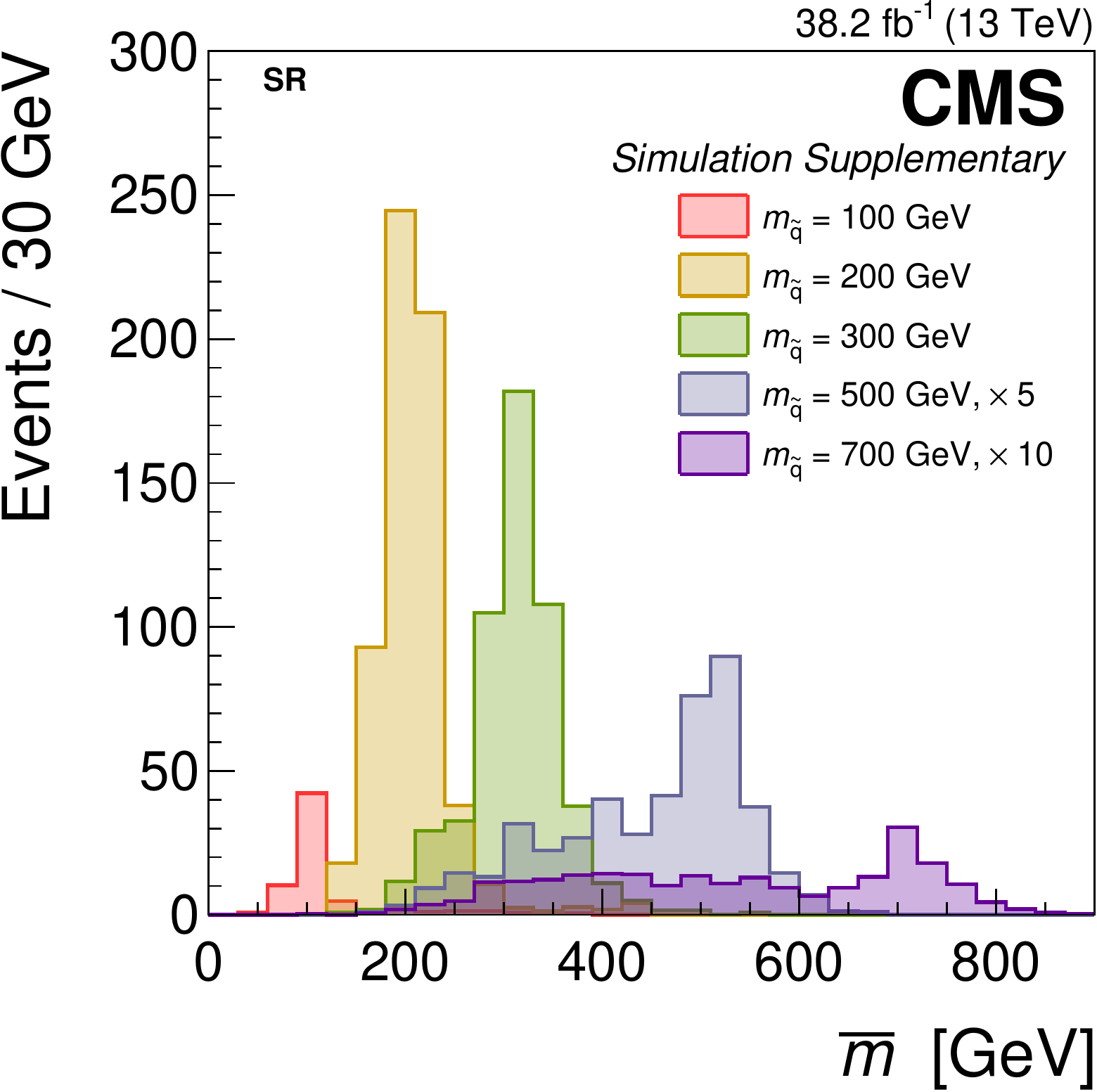}
		\caption{Distributions in \mavg in the SR for a selection of simulated squark signal samples.}
		\label{fig:signals_sq}	
\end{figure}

\begin{figure}[htb!]
	\centering
		\includegraphics[width=0.48\textwidth]{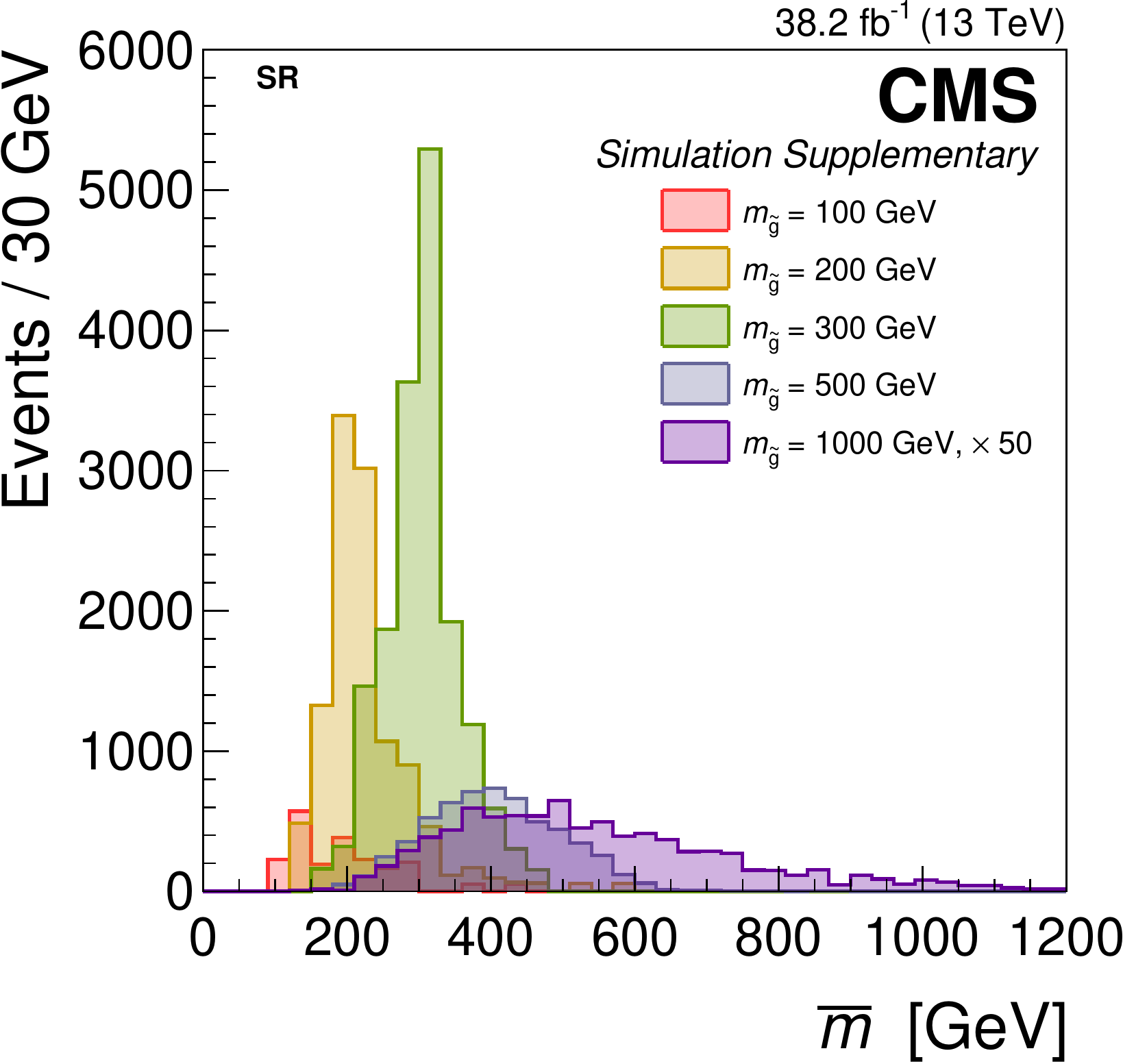}
		\caption{Distributions in \mavg in the SR for a selection of simulated gluino signal samples.}
		\label{fig:signals_sg}	
\end{figure}

\begin{figure}[htb!]
	\centering
		\includegraphics[width=0.48\textwidth]{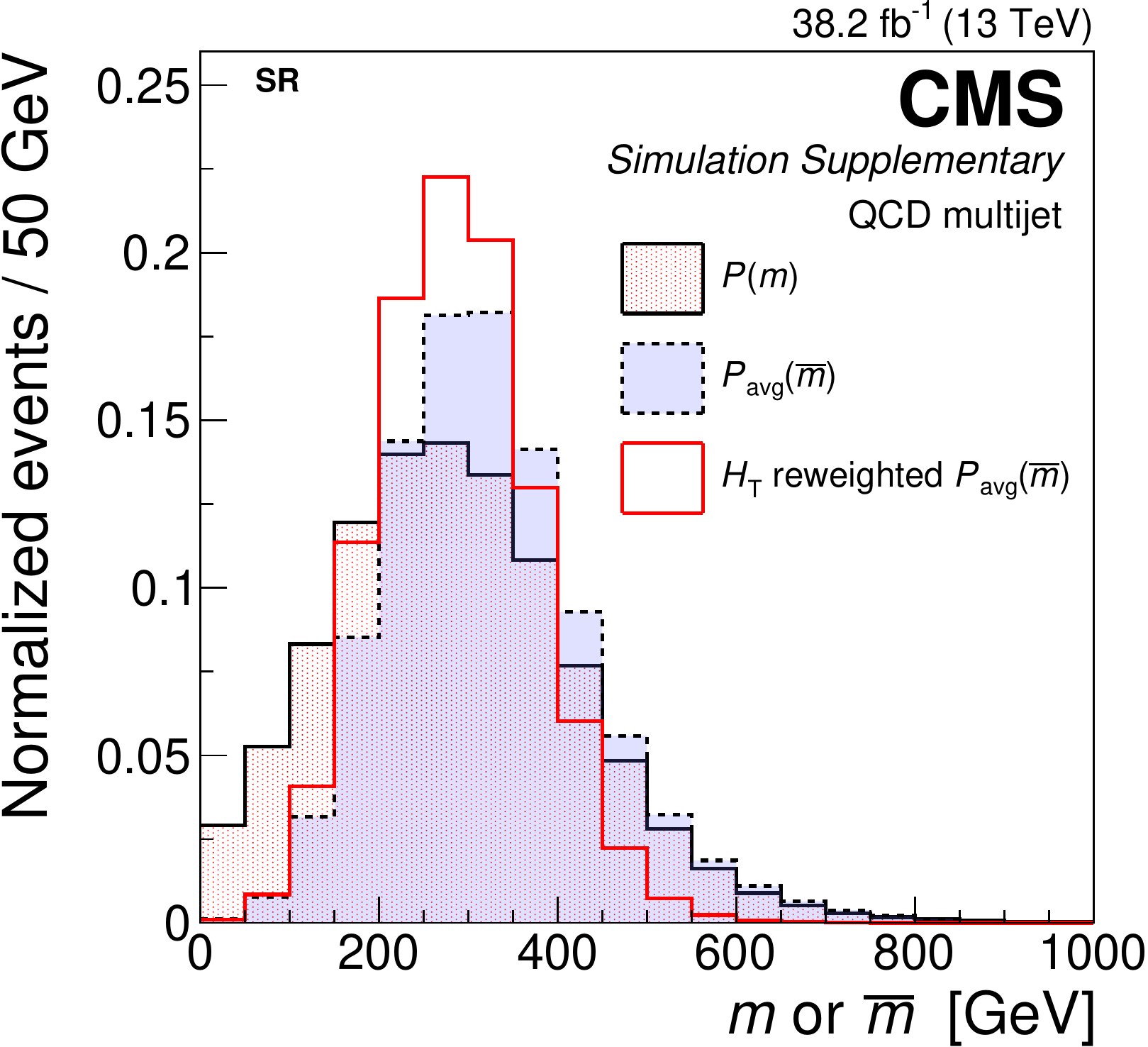}
		\caption{Distributions in \mavg for the tagged \pt-leading fat jets $P(m)$ in the SR and the \mavg PDFs derived from these fat jets with and without \HT reweighting.}
		\label{fig:technique}	
\end{figure}

}

\cleardoublepage \section{The CMS Collaboration \label{app:collab}}\begin{sloppypar}\hyphenpenalty=5000\widowpenalty=500\clubpenalty=5000\vskip\cmsinstskip
\textbf{Yerevan Physics Institute, Yerevan, Armenia}\\*[0pt]
A.M.~Sirunyan, A.~Tumasyan
\vskip\cmsinstskip
\textbf{Institut f\"{u}r Hochenergiephysik, Wien, Austria}\\*[0pt]
W.~Adam, F.~Ambrogi, E.~Asilar, T.~Bergauer, J.~Brandstetter, M.~Dragicevic, J.~Er\"{o}, A.~Escalante~Del~Valle, M.~Flechl, R.~Fr\"{u}hwirth\cmsAuthorMark{1}, V.M.~Ghete, J.~Hrubec, M.~Jeitler\cmsAuthorMark{1}, N.~Krammer, I.~Kr\"{a}tschmer, D.~Liko, T.~Madlener, I.~Mikulec, N.~Rad, H.~Rohringer, J.~Schieck\cmsAuthorMark{1}, R.~Sch\"{o}fbeck, M.~Spanring, D.~Spitzbart, A.~Taurok, W.~Waltenberger, J.~Wittmann, C.-E.~Wulz\cmsAuthorMark{1}, M.~Zarucki
% ...
\vskip\cmsinstskip
% ...
1:  Also at Vienna University of Technology, Vienna, Austria\\
% ...
\end{sloppypar}
\end{document}